\documentclass[journal,comsoc]{IEEEtran}

\usepackage[T1]{fontenc}

\usepackage{amsmath}
\usepackage{overpic}
\usepackage{color}
\usepackage{cite}

\begin{document}

\title{Reconfigurable Intelligent Surfaces: \\ Three Myths and Two Critical Questions}

\author{
\IEEEauthorblockN{Emil Bj{\"o}rnson,  {\"O}zgecan {\"O}zdogan, Erik G. Larsson
\thanks{
\newline\indent The paper was supported by ELLIIT and the Swedish Research Council (VR). 
\newline \indent The authors are with the Department of Electrical Engineering (ISY), Link\"{o}ping University, SE-58183 Link\"{o}ping, Sweden \{emil.bjornson,ozgecan.ozdogan,erik.g.larsson\}@liu.se.}
}}

\maketitle

\begin{abstract}
The search for physical-layer technologies that can  play a key role in beyond-5G systems has started.
One option is reconfigurable intelligent surfaces (RIS), which can collect wireless signals from a transmitter and passively beamform them towards the receiver.
The technology has exciting prospects and is quickly gaining traction in the communication community, but in the current hype we have witnessed how several myths and overstatements are spreading in the literature. 

In this article, we take a neutral look at the RIS technology. We first review the fundamentals and then explain specific features that can be easily misinterpreted. In particular, we debunk three myths: 1) Current network technology can only control the transmitter and receiver, not the environment in between; 2) A better asymptotic array gain is achieved than with conventional beamforming; 3) The pathloss is the same as with anomalous mirrors.

To inspire further research, we conclude 
by identifying two critical questions that must be answered for RIS to become a successful technology: 1) What is a convincing use case for RIS?; 2) How can we estimate channels and control an RIS in real time?
\end{abstract}

\IEEEpeerreviewmaketitle

\section*{Introduction}

The electromagnetic waves that carry information in wireless communications interact with objects and surfaces on their way from the transmitter to the receiver.
Although the superposition of many propagation paths gives rise to random-like fading phenomena, every propagation path has a constant behavior.
However, there exist engineered materials whose interactions with electromagnetic waves are not constant but reconfigurable.
These materials are not naturally occurring but can be manufactured and deployed to shape the propagation environment.
The prospects of including such \emph{reconfigurable intelligent surfaces (RIS)} as a part of  beyond-5G network architectures are attracting much attention \cite{Huang2019,Renzo2020a}. RIS have  also been called \emph{software-controlled metasurfaces} \cite{Liaskos2018a} and \emph{intelligent reflecting surfaces} \cite{Wu2019a}.

\begin{figure*}[t!]
	\centering 
	\begin{overpic}[width=2.05\columnwidth,tics=10]{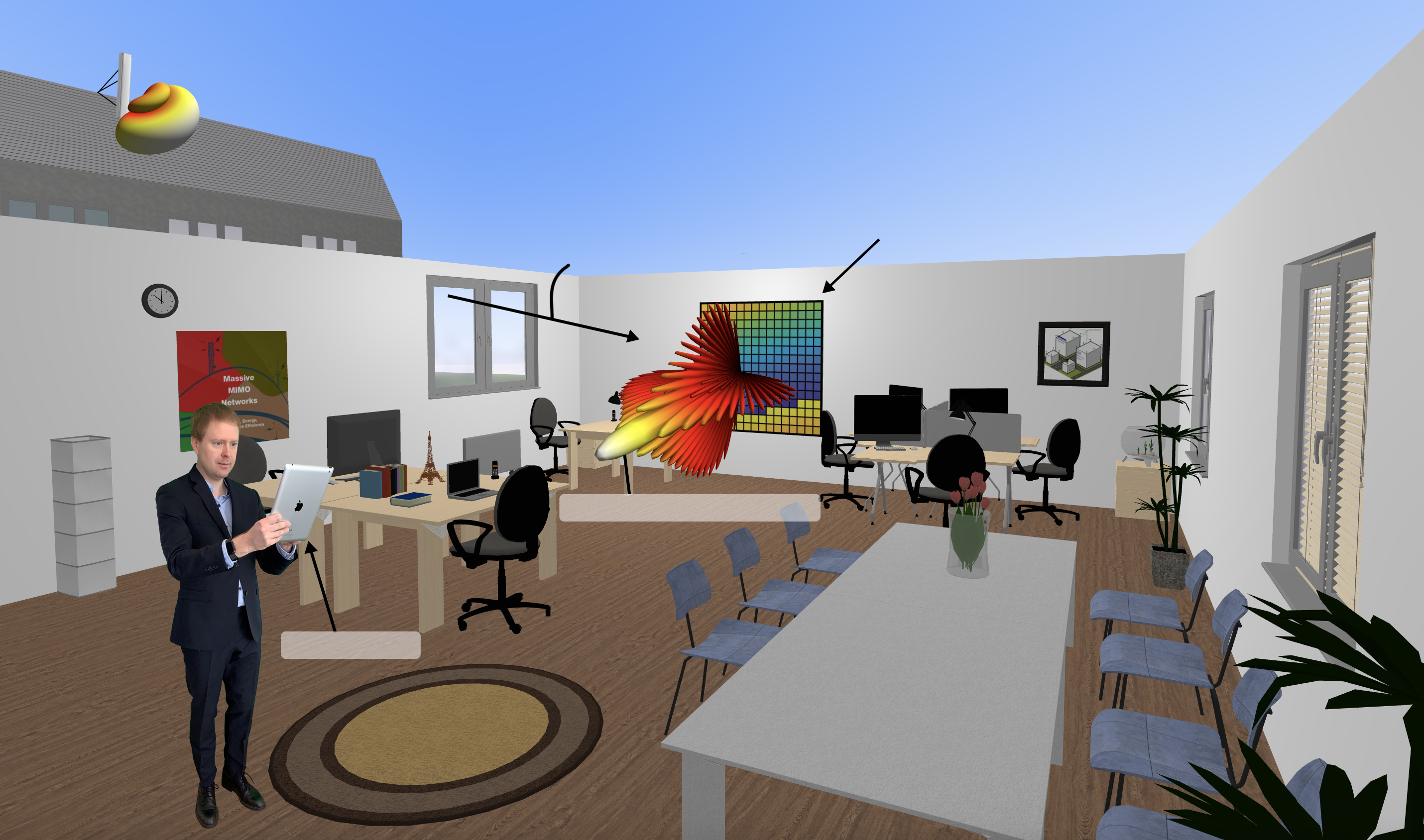}
	\put(8,56.5){1) Transmitter}
	\put(37,41){2) Signal reaching the RIS}
	\put(62,47){3) RIS with elements}
	\put(62,45){configured to shape a beam}
	\put(62,43){towards the receiver}
	\put(39.5,22.8){4) Beam from the RIS}
	\put(20,13.2){5) Receiver}
\end{overpic} 
	\caption{A typical use case of an RIS, where it receives a signal from the transmitter and re-radiates it focused towards the receiver. To focus the beam in the right direction, the RIS must be configured properly.}
	\label{figureBasicExample} 
\end{figure*}
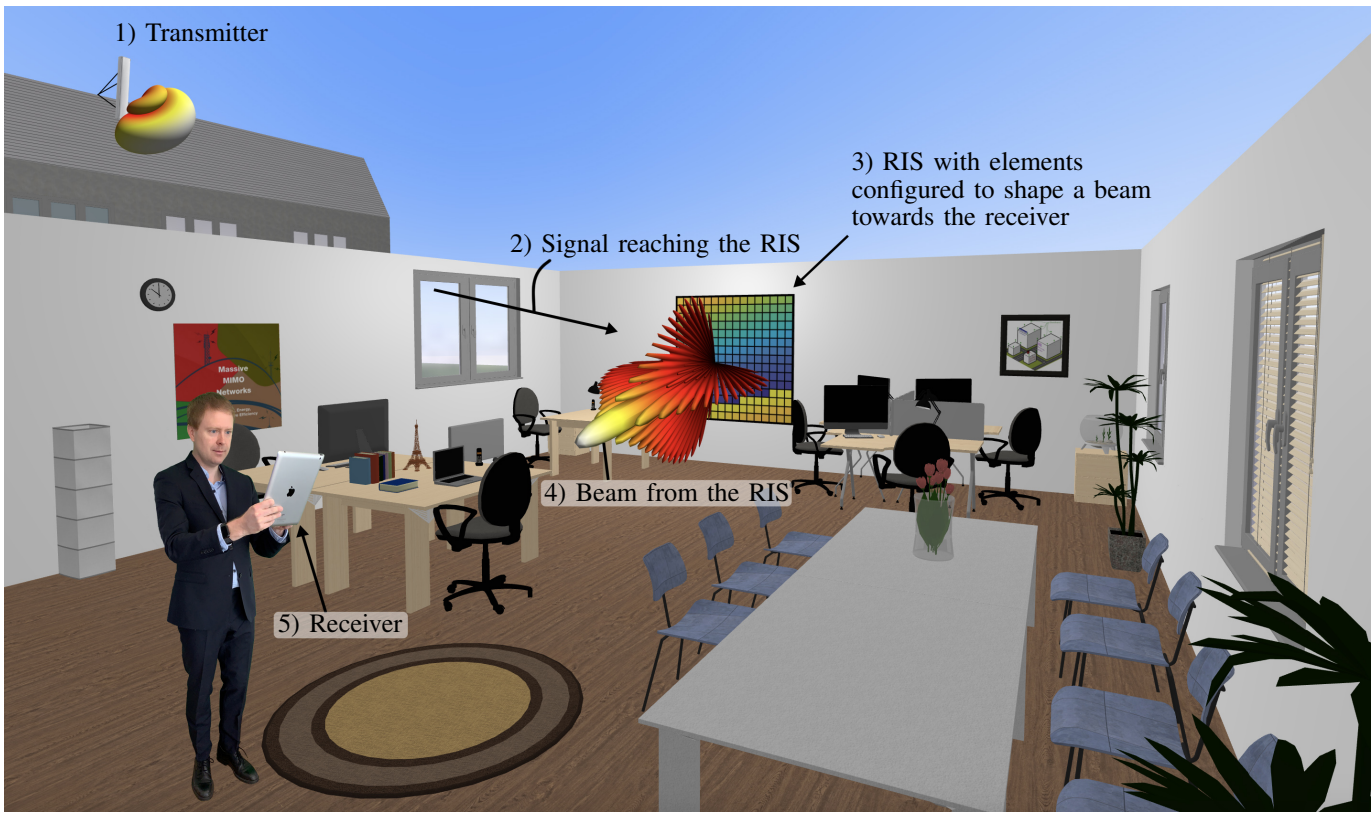

A basic use case of RIS is illustrated in Fig.~\ref{figureBasicExample}, where a rooftop-mounted base station (BS) is transmitting to an indoor user.
Suppose the building materials are such that the direct path through the wall experiences massive penetration losses, while the path through the window only experiences minor losses.
Inside the window, an RIS is deployed to capture signal energy proportional to its area and re-radiate it in the shape of a beam towards the receiver.
To ensure the beam is focused towards the user device, wherever it is in the room, the RIS must be reconfigurable.
By using an RIS in this setup, the signal-to-noise ratio (SNR) can be improved.

An RIS is a thin surface composed of $N$ elements, each being a reconfigurable scatterer: a small antenna that receives and re-radiates without amplification, but with a configurable time-delay \cite{Liaskos2018a}.
For narrowband signals, this delay corresponds to a phase-shift.
Assuming the phase-shifts are properly adjusted, the $N$ scattered waves will add constructively at the receiver. This principle resembles traditional beamforming: each element has a fixed radiation pattern but the collection of phase-shifts determines where constructive interference among the scattered waves occurs. The color pattern at the RIS in Fig.~\ref{figureBasicExample} represents the phase-shifts necessary to steer a beam towards the receiver. Each element is substantially smaller than the wavelength (e.g., a fifth of the wavelength in each direction \cite{Tsilipakos2020a}) so it scatters signals almost uniformly, giving the surface the ability to form equally strong beams in any direction \cite{Ozdogan2019a}.

The propagation analysis of an RIS essentially entails (i) finding the Green's function of the signal source (a sum of spherical waves if close, or a plane wave if far away), (ii) computing the impinging field at each RIS element, (iii) integrating this field over the surface of each element to find the current density, (iv) computing the radiated field from each element, and (v) applying the superposition principle to find the field at the receiver. Since the elements are small, one can approximate the re-radiated field by pretending each element is a point source and then the received signal is a superposition of phase-shifted, amplitude-scaled source signals \cite{Ozdogan2019a}.

There are many prospective use cases for RIS-aided wireless communications, in addition to improving the SNR as in Fig.~\ref{figureBasicExample}. The RIS can also mitigate interference between users that are spatially multiplexed or limit the signal-leakage outside the intended coverage area, to mitigate eavesdropping \cite{Liaskos2018a,Wu2019a,Renzo2020a}. Support for wireless power transfer, backscattering, and spatial modulation is also conceivable; most things that can be implemented using antenna arrays can also be carried out by an RIS \cite{Tsilipakos2020a}.

The definition of an RIS is a surface with real-time reconfigurable scattering properties (e.g., amplitude, delay, and polarization)  that is controlled to improve the communication performance. The concept is often connected with metasurfaces, which are two-dimensional surfaces consisting of arrays of reconfigurable elements of metamaterial \cite{Tsilipakos2020a}. However, there are other potential ways of implementing RIS \cite{Renzo2020a}. One example is using small patch antennas terminated with an adjustable impedance. In any case, the reconfigurability will likely be limited to a finite set of states per element (with given delays and amplitudes) and mutual coupling between adjacent elements is another limitation.
We refer to \cite{Tsilipakos2020a} for a recent survey on the implementation aspects for metasurfaces.
There are decades of research on reflectarrays and array lenses \cite{Hum2014a}, which are architectures for building transmitters consisting of a feed antenna that sends the signal via a reconfigurable surface capable of electronically tunable beamforming. The key difference is that an RIS is not co-located with the transmitter or receiver, but deployed in between to aid the communication.

\section*{Basic Features and Related Myths}

We will now describe three fundamental features that the RIS technology possess. Along the way, we will also debunk three myths that are flourishing in the literature.

\subsection*{Feature 1: Creating Controllable Radio Environments}

A key feature of RIS is the ability to alter how wireless signals propagate between the transmitter and receiver.
It is a technology for creating controllable/smart/programmable radio environments, which are defined as environments that can customize how signals propagate from the transmitter to the receiver \cite{Renzo2020a}.
This feature enables joint optimization of the transmitter/receiver and the controllable entities in the environment, using channel state information (CSI).
When motivating the novelty of this feature, the following claim has been made repeatedly \cite{Liaskos2018a,Wu2019a,Basar2019a}.

\textit{Myth 1: Current network technology can only control the transmitter and receiver, not the environment in between.}

Many wireless systems indeed consist of a transmitter that communicates with a receiver without the involvement of other entities.
The radio environment is then uncontrollable; the transmitter and receiver must conform to it by adaptive modulation/coding, beamforming, and power control.
However, this is a choice made by the network provider because the technology for controlling the signal propagation between the end points has existed all along. The wireless repeater was invented in 1899 and advanced relaying technology, capable of adaptively improving the channel between the transmitter and receiver, has been supported by cellular standards since 3G \cite{Dohler2010a}. Hence, the statement above is a myth.

We will now put the RIS technology into a historical context.
The term \emph{cooperative communications} is broadly used to refer to network architectures containing entities between the transmitter and receiver that enhance the physical channel, by exploiting diversity, beamforming, and/or multiplexing gains \cite{Dohler2010a}. These entities are co-optimized with the transmitter and receiver, thus satisfying the definition of controllable radio environments.
Two main categories are \emph{transparent relaying} and \emph{regenerative relaying}.
In the former category, each relay is an entity that receives a signal from the transmitter and processes it in analog (or digitally) before re-radiating it towards the receiver. Amplify-and-forward is a classic protocol for creating additional signal paths by re-radiating an amplified signal in a way that can be transparent to the receiver. No baseband processing is required, only amplification. In regenerative relaying, each relay is decoding the received signal and processes it in the digital baseband, before retransmitting it in an optimized manner towards the receiver. Decode-and-forward (DF) is a common example. Classical relays operate in half-duplex, where reception and retransmission are separated in time, but regenerative full-duplex relays capable of receiving and transmitting simultaneously are emerging \cite{Heino2015a}.

The RIS technology is unique in that it fills an empty slot in the relaying taxonomy: it is a transparent relay with a full-duplex protocol \cite{Huang2019}, thus it affects the propagation in real-time.
The potential advantage over traditional relays is that large surfaces can be implemented with reduced energy consumption and cost since the use of printed metamaterial requires no amplifiers, but only power dissipation in the hardware controlling the reconfigurability. The drawback is the reduced signal range due to the lack of amplification.

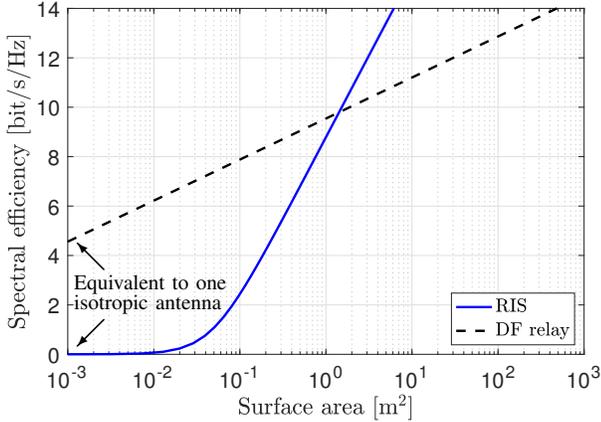
\begin{figure}[t!]
	\centering
	\begin{overpic}[width=\columnwidth,tics=10]{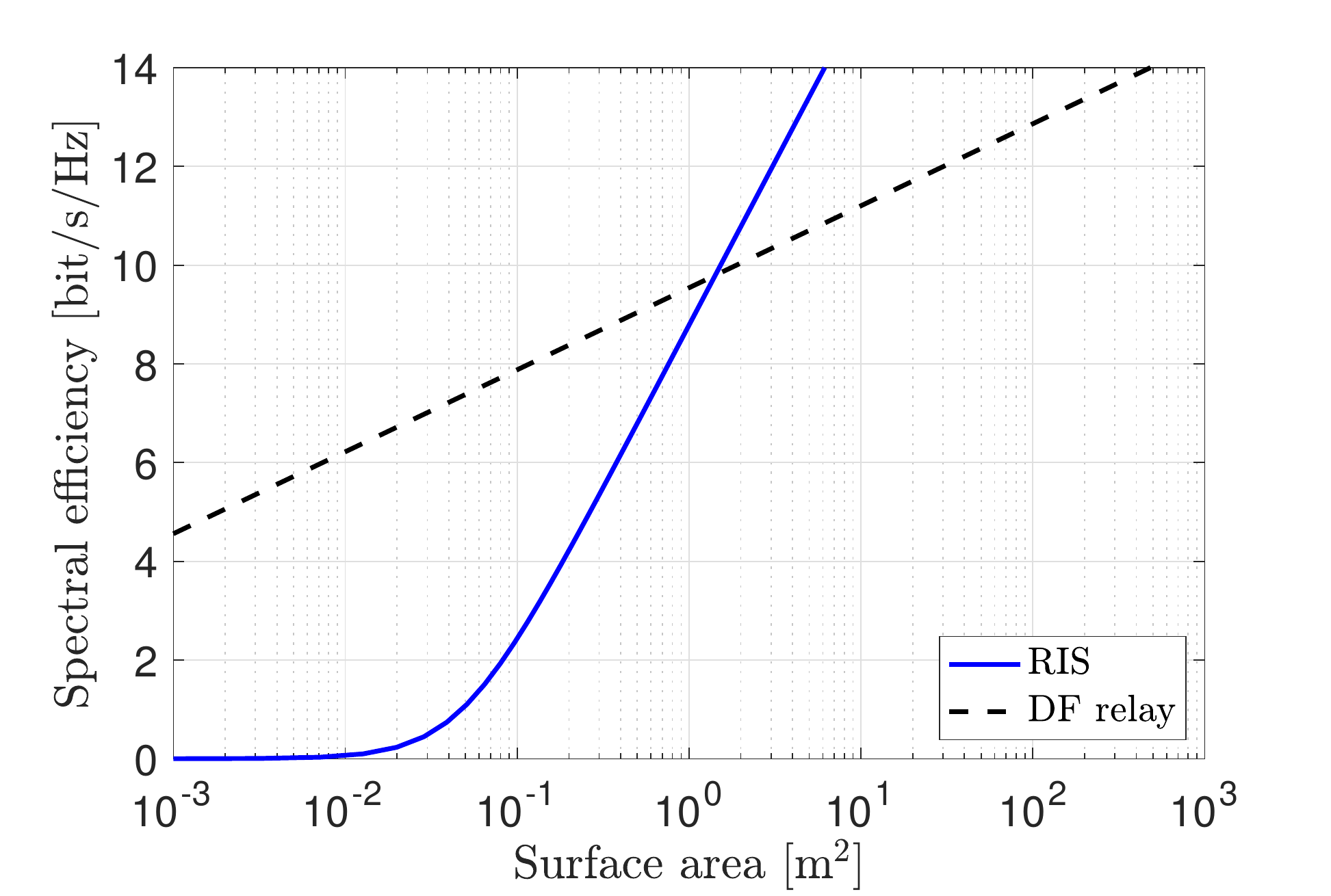}
\put(14,20){\footnotesize Equivalent to one}
\put(14,17){\footnotesize isotropic antenna}
\put(18.5,15.5){\vector(-1,-1){4}}
\put(18.5,23){\vector(-1,1){4}}
\end{overpic}  \vspace{-6mm}
	\caption{This figure revisits the setup in Fig.~\ref{figureBasicExample} and compares the use of an RIS with the use of a multi-antenna DF relay deployed at the same place. The direct path is assumed non-existing, while there are line-of-sight paths via the RIS/relay. The transmitter is 300\,m from the RIS/relay, while the user is 10\,m from it. The transmit power is equivalent to 10 W per 20 MHz at the BS and 0.1 W at the relay. The antenna gain is 10 dBi at the BS and 0 dBi at the relay and user. The glass penetration loss is $-20$ dB and the noise figure is $10$ dB.
The figure shows the SE achieved by an RIS and a DF relay for different surface areas.
	For practical SEs (below 8 bit/s/Hz), the DF relay is smaller, while the advantage of the RIS is the lack of power amplifiers and full-duplex mode.}  \vspace{-2mm}
	\label{figureMyth1} 
\end{figure}

Fig.~\ref{figureMyth1} illustrates this in a setup that resembles the one illustrated in Fig.~\ref{figureBasicExample}, but where the RIS is possibly replaced by a multi-antenna half-duplex repetition-coded DF relay (a simple but suboptimal relaying scheme). Perfect CSI is assumed and each RIS element scatters all the incoming signal energy with a perfectly controlled phase.
The figure shows the array's surface area required to achieve a particular spectral efficiency (SE) when using either an RIS or a DF relay. The results are frequency-independent but the number of elements that fits into the surface area grows quadratically with the wavelength.
We observe that the DF relay can have a much smaller form factor than the RIS, except if very high SE is required.
The reason is that the DF relay achieves a much higher SNR but it also needs a higher SNR to achieve the same SE since it operates in half-duplex,
whereas the RIS operates in full-duplex.

In summary, the RIS technology can control/optimize the propagation environment between the transmitter and receiver, just as previous relaying technologies. The unique feature of RIS is that it reduces the hardware complexity at the price of requiring a larger surface.

\subsection*{Feature 2: Passive Beamforming}

Beamforming appears when delayed copies of the same signal are emitted from multiple antennas. This gives rise to constructive interference at spatial locations where the copies are received synchronously and destructive interference elsewhere.
If the time-delays at $N$ transmit antennas are tuned to achieve constructive interference at the receiver, it will receive $N$ times more power than if the same total power was transmitted from a single antenna.
This is the conventional \emph{array gain} that shows how the beamformed signal becomes more spatially focused as the array size grows.

An RIS is capable of passive beamforming. It receives signal power from the transmitter proportional to its surface area, which is proportional to the number of elements, $N$.
When the RIS re-radiates the signal, with time-delays selected to beamform at the receiver, an array gain of $N$ is obtained just as with conventional beamforming.
The combination of these two effects, both being proportional to $N$, leads to an SNR at the receiver proportional to $N^2$. This is called the ``square law'' \cite{Wu2019a}.

Suppose we compare the setup in Fig.~\ref{figureBasicExample} with the case when the RIS is replaced with a transmitter array having the same size. The RIS setup will achieve an SNR that grows as $N^2$, while the SNR in the latter setup only grows as $N$. It has been claimed that these are asymptotic scaling laws \cite{Wu2019a}, which means that the SNR with the RIS grows unboundedly with $N$ at the order of $N^2$ and eventually becomes larger than with the transmitter array. This is incorrect.

\textit{Myth 2: A better asymptotic array gain is achieved than with conventional beamforming.}

The first issue with this statement is that array gains of the type described above only appear when the surface area (of the RIS or transmitter array) is small compared to the propagation distances.
The transmitter/receiver must be in the geometric far-field of the surface so that the pathloss is approximately the same to all parts of the surface.
Since the surface area grows with $N$, the far-field approximation eventually breaks down as $N$ increases.
 Neither linear nor quadratic asymptotic power scaling laws can exist 
since the law of conservation of energy dictates that we cannot receive more energy than was transmitted. Nevertheless, the SNR achieved with an RIS actually grows quadratically with the number of elements for many practically-sized surfaces \cite{Bjornson2020a}.
 Hence, it might seem possible that a better SNR can be achieved in the RIS setup when considering large, equal-sized arrays. The second issue with the statement is the premise that the quadratic power scaling is advantageous.
 The pathloss from the transmitter to each RIS element is huge in the far-field, thus it is more accurate to say that the power loss between the transmitter and RIS reduces as $1/N$ \cite{Bjornson2020a}.
 The SNR achieved with an RIS cannot surpass the SNR achieved when replacing it by an equal-sized antenna array transmitting with the same power as in the RIS case, but the difference reduces as $1/N$.

\begin{figure}[t!]
	\centering
	\begin{overpic}[width=\columnwidth,tics=10]{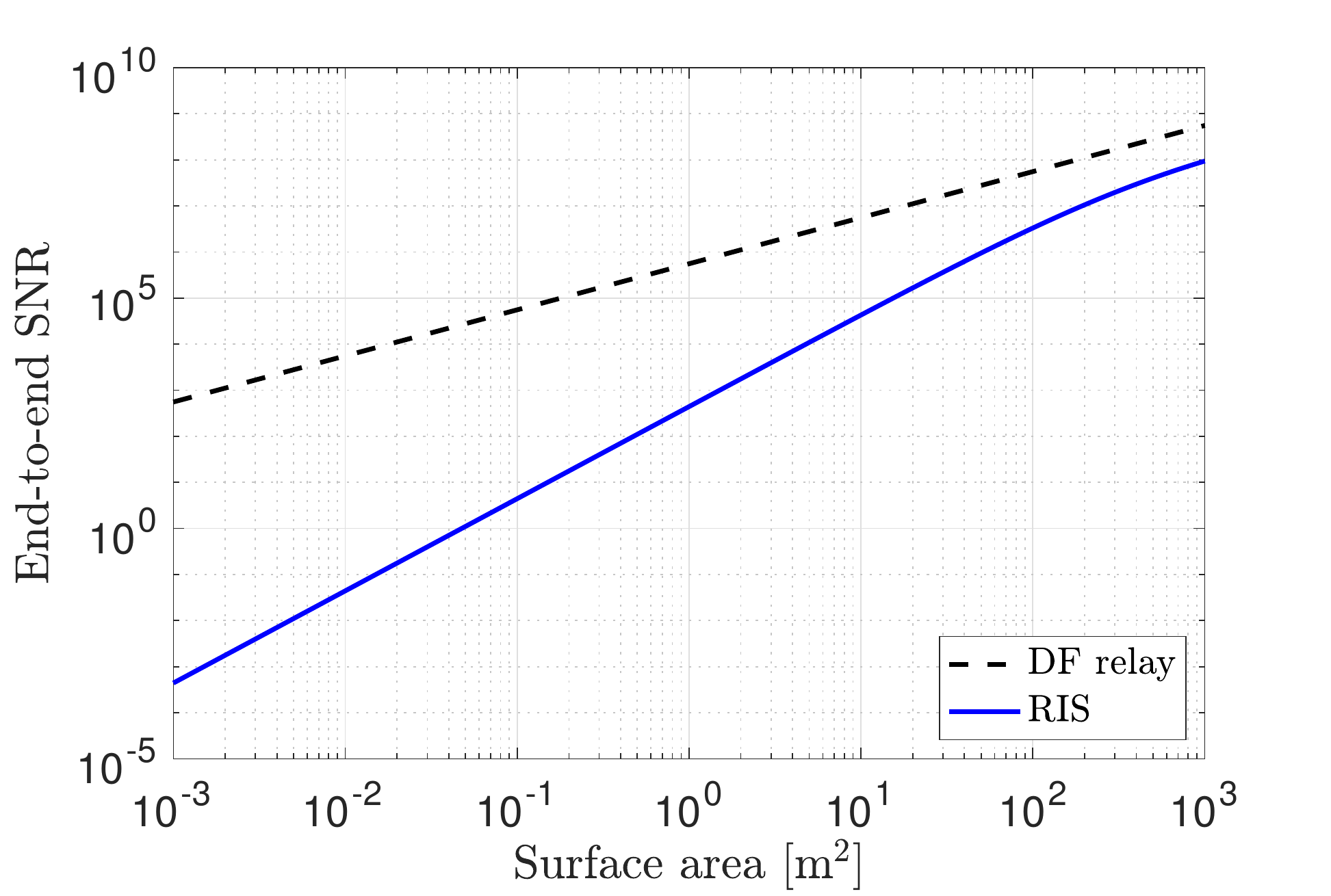}
\put(66,39){\footnotesize The ``square law''}
\put(66,36){\footnotesize  tapers off}
\put(82,43){\vector(0,1){7}}
\end{overpic} 
	\caption{This figure continues the example from Fig.~\ref{figureMyth1} by comparing the end-to-end SNR achieved by the RIS and the DF relay. The RIS benefits from the ``square law'' by achieving a steeper slope for practical surface areas (below 100 m$^2$). Nevertheless, the DF relay is consistently providing a better SNR and both curves converge to a finite number since there are no asymptotic scaling laws.}
	\label{figureMyth2} 
\end{figure}

The reason that the DF relay outperformed the RIS for most SE values in Fig.~\ref{figureMyth1} is that the RIS suffers from the power loss  inherent in the ``square law''.
To demonstrate this, Fig.~\ref{figureMyth2} revisits the example by showing the end-to-end SNR achieved with the RIS and DF relay for different surface areas. Since we use logarithmic scales, the quadratic array gain is observed from the steeper slope of the RIS curve. However, this curve begins at a much smaller value and when it approaches the DF relay curve, the steeper slope has tapered off.
Both curves will eventually converge to a finite number \cite{Bjornson2020a}.
The reason that the RIS became preferable for very high SEs in Fig.~\ref{figureMyth1} is that the SNR gap eventually becomes so small that the half-duplex operation of the DF relay becomes the bottleneck.

In summary, an RIS is capable of passively beamforming a signal towards the receiver. Due to the faster-than-linear SNR-scaling, physically large surfaces are highly preferable.

\subsection*{Feature 3: Synthesizing a Different Surface Shape}

The RIS can not only form a beam, it can synthesize the scattering behavior of an arbitrarily-shaped surface of the same size. For example, it can create a superposition of multiple beams or act as a diffuse scatterer \cite{Tsilipakos2020a}.

\begin{figure*}[t!]
	\centering 
	\begin{overpic}[width=1.8\columnwidth,tics=10]{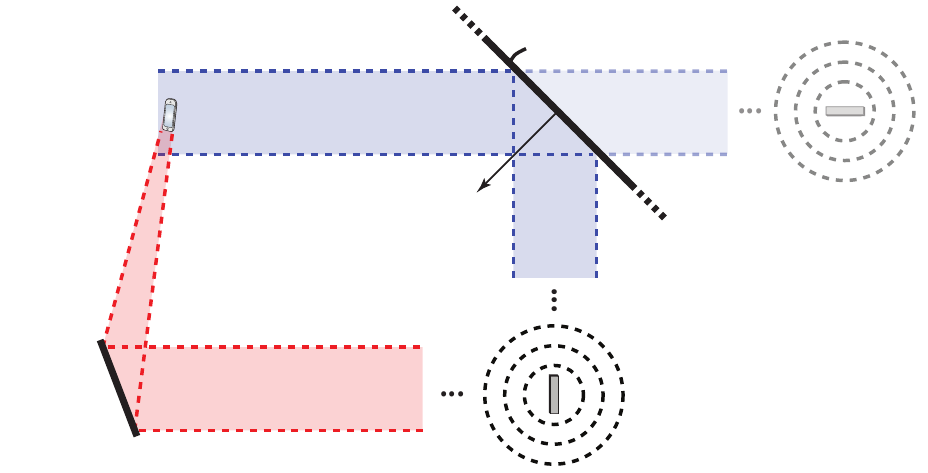}
	\put(1,8){Finite-sized}
	\put(4.5,5.5){RIS}
	\put(58,45){Infinite-sized mirror}
	\put(85,29){Mirror image}
	\put(85,26.5){of transmitter}
	\put(45,31.5){Normal}
	\put(69,8){Transmitter}
	\put(8,38){Receiver}
	\put(65.5,24){Impinging}
	\put(65.5,21.5){plane wave}
	\put(24,44.2){Reradiated plane wave}
	\put(1.5,24){Reradiated}
	\put(1.5,21.5){signal beam}
	\put(22,14.5){Impinging plane wave}
\end{overpic}  \vspace{-3mm}
	\caption{A mirror reflects an impinging plane wave as a plane wave in an angular direction determined by the law of reflection, so the receiver perceives the transmitter as being located at the mirror image location. An RIS can both configure the angle of the reflected beam and its shape, thus it should not be interpreted as an anomalous mirror. The figure illustrates how the RIS focuses the signal at the receiver to maximize the SNR.}
	\label{figureReflectionExample} 
\end{figure*}

A common example is to synthesize an \emph{anomalous mirror/reflector}. A mirror is a surface that reflects an impinging plane wave as an outgoing plane wave, also known as specular reflection.
A conventional mirror satisfies the law of reflection: the angles of the impinging and reflected waves to the surface normal are the same but on opposite sides, as illustrated by the blue ribbons in Fig.~\ref{figureReflectionExample}.
An anomalous mirror reflects impinging plane waves as outgoing 
plane waves with a different ``unnatural'' angle to the surface normal \cite{Renzo2020a}. A conventional mirror is an infinitely large homogeneous surface and approximations thereof appear naturally (e.g., a metal plate or water surface). In contrast, an anomalous mirror is not naturally appearing but can be synthesized by an engineered inhomogeneous surface.
A property of mirrors is that the receiver observes the transmitting source as being behind the mirror.
One can analyze the wave propagation as if the transmitter is moved to the location of the mirror image, as illustrated in Fig.~\ref{figureReflectionExample}.

It has been stated that an RIS can generally be viewed as an anomalous mirror if it has a width and length larger than ten wavelengths \cite{Basar2019a}. If that is the case, the pathloss in Fig.~\ref{figureReflectionExample}  can be computed based on the sum of the distance from the transmitter to the RIS and from the RIS to the receiver \cite{Basar2019a}, which is the distance from the mirror image to the receiver. These are myths that are summarized as follows.

\textit{Myth 3: The pathloss is the same as with anomalous mirrors.}

An ideal mirror reflects a signal with zero beamwidth. If a plane wave is impinging on a finite-sized RIS that is configured to focus the signal towards a receiver located in the far-field, then the radiated field will be strongest in the angular direction of the receiver but it will not be a plane wave. Far-field focusing is called beamforming and the beamwidth is the same as for beamforming from an equal-sized transmitter array. 
Hence, the half-power beamwidth of the reflected signal is inversely proportional to the size of the RIS (measured in wavelengths) and becomes $6^\circ$ for a surface that is ten wavelengths in each dimension \cite{Ozdogan2019a}. 

Mirrors and plane waves are theoretical idealizations that 
only appear approximately in practice. They can be fairly accurate approximations in visible light and are, thus, used in geometrical optics to analyze imaging.
The situation is different in the radio spectrum used for communications.
A surface that our eyes perceive as a mirror might be far from mirror-like for radio signals.
Since the wavelength is roughly 100000 times larger in radio spectrum than in visible light (e.g., comparing green light at 600 THz with a radio signal at 6 GHz), a surface must be 100000 times larger in each dimension to identically reflect signals.
The transmitter must be 100000 times further away if its emitted spherical waves should be approximated as planar, and the receiver must be 100000 times further away to perceive the reflected signals as plane waves.
Since mirrors only exist in asymptotic limits, there are no finite-sized surfaces that always can be approximated as mirrors. If the RIS is viewed from far enough away, its radiated field will have a beamwidth that is inversely proportional to its size.

Even if we limit the scope to setups where conventional mirrors approximately exist, the statement above remains a myth since the pathloss achieved by an equal-sized RIS is widely different. An RIS can both affect the direction and shape of the reflected signal \cite{Bjornson2020a}, as illustrated by the red ribbons in Fig.~\ref{figureReflectionExample} where the signal is focused at the receiver. For this reason, the SNR achieved by the RIS is proportional to $N^2$ and is inversely proportional to the product of the squared distances to the RIS \cite{Ozdogan2019a}, rather than inversely proportional to the squared sum of the distances as with a mirror.

\begin{figure}[t!]
	\centering
	\begin{overpic}[width=\columnwidth,tics=10]{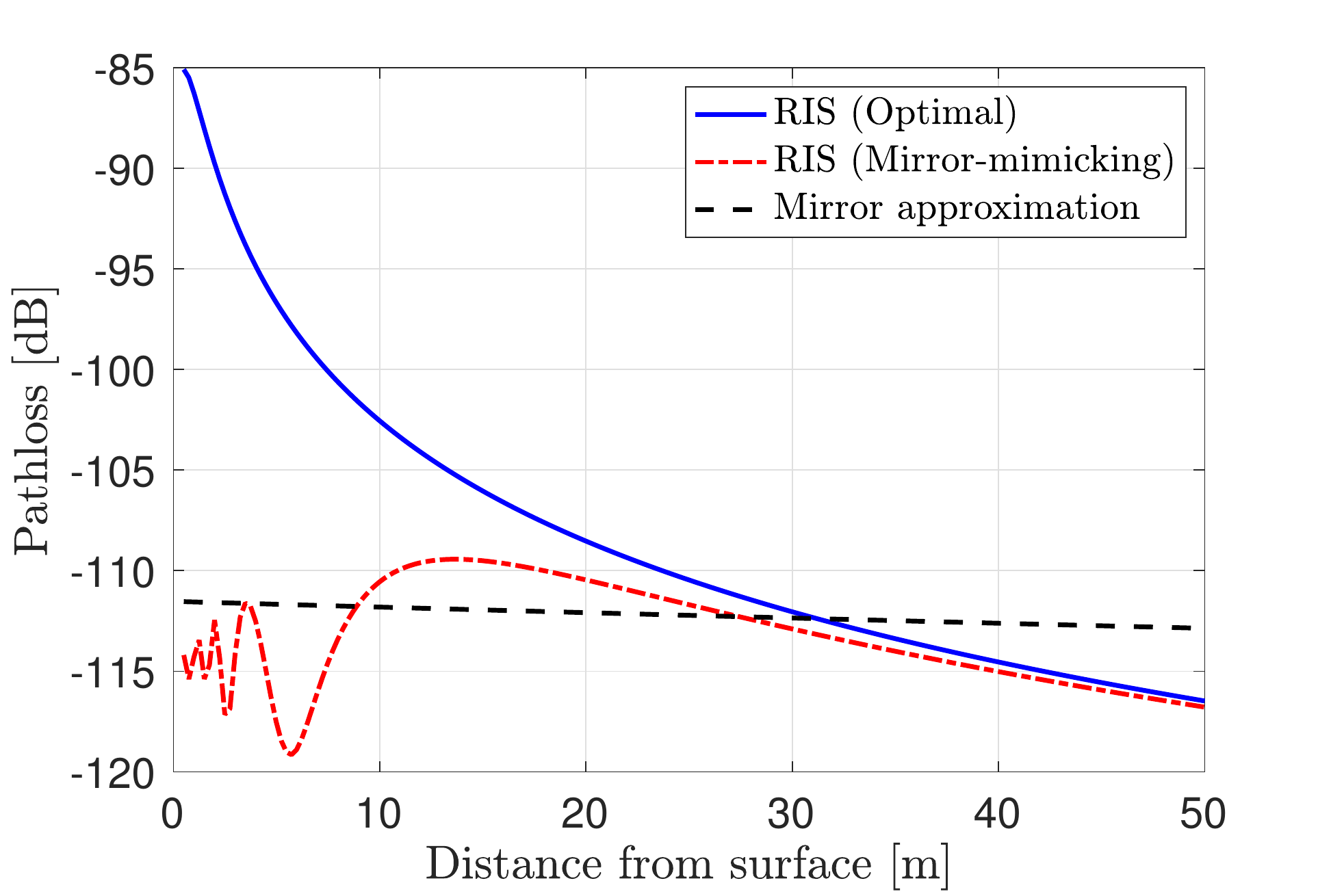}
\put(43,34){Only point where the mirror}
\put(43,30){approximation is accurate}
\put(61,28){\vector(0,-1){5}}
\end{overpic} 
	\caption{This figure revisits the setup from the previous figures and considers an RIS that is $2 \times2$\,m, which represents $20 \times 20$ wavelengths at a 3 GHz frequency and $200 \times 200$ wavelengths at a 30 GHz frequency. The figure shows the end-to-end pathloss as a function of the distance between the RIS and the receiver. An optimally configured RIS is compared with an RIS that is configured to mimic a mirror and the pathloss obtained if it was an ideal mirror. It is clear that an RIS can generally not been interpreted as a mirror.}
	\label{figureMyth3} 
\end{figure}

To explain the fundamental differences, Fig.~\ref{figureMyth3} continues the example from Figs.~\ref{figureMyth1} and \ref{figureMyth2} by showing how the end-to-end pathloss depends on how far the receiver is from the RIS (the distance between the transmitter and RIS is as before).
The solid curve is for an RIS that is optimized to achieve the highest SNR, while the dashed curve represents an anomalous mirror.
We notice that a mirror is a poor approximation of an RIS at most distances.
When the receiver is far from the RIS, the pathloss is worse than with a mirror since the RIS is too small to emit approximately plane waves.
When the receiver is close to the RIS, the pathloss is instead much better than with a mirror.
This is like when you look into a large mirror and your reflection only appears in a small part of it; the rest of the mirror is not needed.
A well-configured RIS makes use of the entire surface by focusing the signal at the receiver as illustrated in Fig.~\ref{figureReflectionExample}.
The dash-dotted curve in Fig.~\ref{figureMyth3}  represents a mirror-mimicking RIS that is configured to delay the signals as a cutout from an infinitely large anomalous mirror would do. This curve is close to the optimized RIS when the receiver is far from the surface, while it begins to oscillate in the vicinity of the mirror approximation at shorter distances. This indicates that the mirror analogy can be used for identifying suitable time-delays when the transmitter and receiver are far away.  This is the reason why (approximate) anomalous reflection is a canonical use case in the electromagnetic literature \cite{Tsilipakos2020a}, where pathloss modeling is not considered.

In summary, the pathloss achieved with an RIS does not coincide with that of an anomalous mirror. When the receiver is far from the surface, it is too small to behave like a mirror. When the receiver is near the surface, the RIS can approximate the mirror behavior but it would be suboptimal; a mirror beamforms to points infinitely far away, while a RIS can focus on the actual receiver location.
One way to describe the capabilities of an RIS is as a parabolic reflector with curvature and direction that can be electronically steered, but that is also a simplification since an RIS is capable of mimicking the scattering of arbitrarily-shaped objects having the same size.

\section*{Critical question 1: What is a convincing use case for RIS?}

An immense amount of time and resources are required to bring a new technology concept, such as RIS, from theory to practice.
Very convincing benefits compared to existing technologies must be established to motivate such an investment; we essentially need to demonstrate 10 times improvements with respect to a practically important performance metric, not just 20\% gains that might disappear in an imperfect implementation.
Massive MIMO (multiple-input multiple-output) and mmWave communications passed this test in the 5G development since the former can increase the number simultaneously served users by ten times while the latter can increase the data rate per user by ten times  using much wider bandwidths.
Several other ``5G-branded'' technologies failed the test because the gains were too limited.

RIS technology has many technical features beyond current mainstream technology \cite{Renzo2020a,Tsilipakos2020a}.
However, to motivate the practical development of RIS technology, the critical question is: \emph{what is a convincing use case?}
The question is open; RIS is a hammer looking for a nail.
There is no shortage of visions on what an RIS can be used for (some ideas were listed in the introduction) but will it excel at anything?
Coverage extension is one option but Fig.~\ref{figureMyth1} showed that conventional half-duplex relaying is a competitive solution, and full-duplex regenerative relays are emerging \cite{Heino2015a}.
Since each RIS element must be identically configured over the entire frequency band, the RIS technology has a further competitive disadvantage over wideband channels.
Improved spatial multiplexing and interference mitigation is another potential use case, but then it needs to beat  Cell-free Massive MIMO, which is the emerging deployment of distributed jointly-operating antennas.
Perhaps it is in terahertz bands, where the implementation of coherent transceivers is truly challenging and the sparse channels make additional propagation paths useful even if they are weak, that the RIS technology will be most beneficial. These are just speculations since there is no hard evidence yet.

\section*{Critical question 2: How can we estimate channels and control an RIS in real time?}

The envisioned use cases of RIS critically depend on a proper configuration of the elements based on CSI.
There are two reasons why channel acquisition is particularly challenging with RIS.
Firstly, unlike conventional transceiver architectures, an RIS is not inherently equipped with transceiver chains. It lacks sensing capabilities but simply ``reflects'' the impinging signals. Therefore, conventional channel estimation methods cannot be utilized.
Secondly, introducing an RIS into an existing setup will increase the number of channel coefficients proportionally to the number of elements, $N$.
As shown earlier, a large $N$ is needed for RIS to be competitive, thus the estimation overhead might be huge. A key question is: \emph{can an RIS be real-time reconfigured to manage user mobility?}

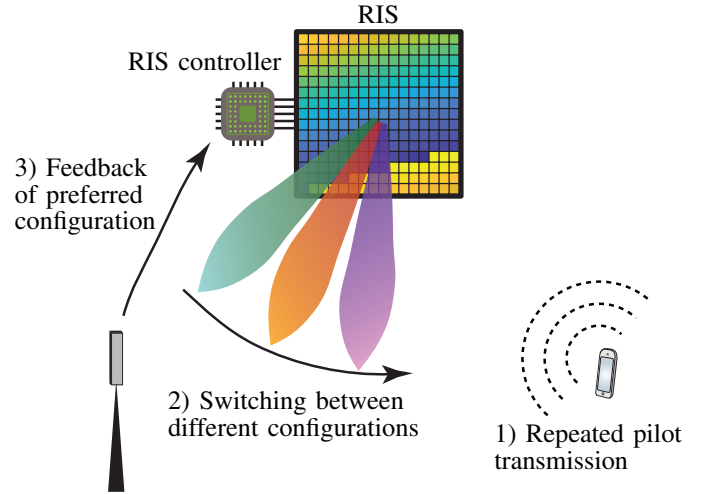
\begin{figure}[t!]
	\centering
	\begin{overpic}[width=\columnwidth,tics=10]{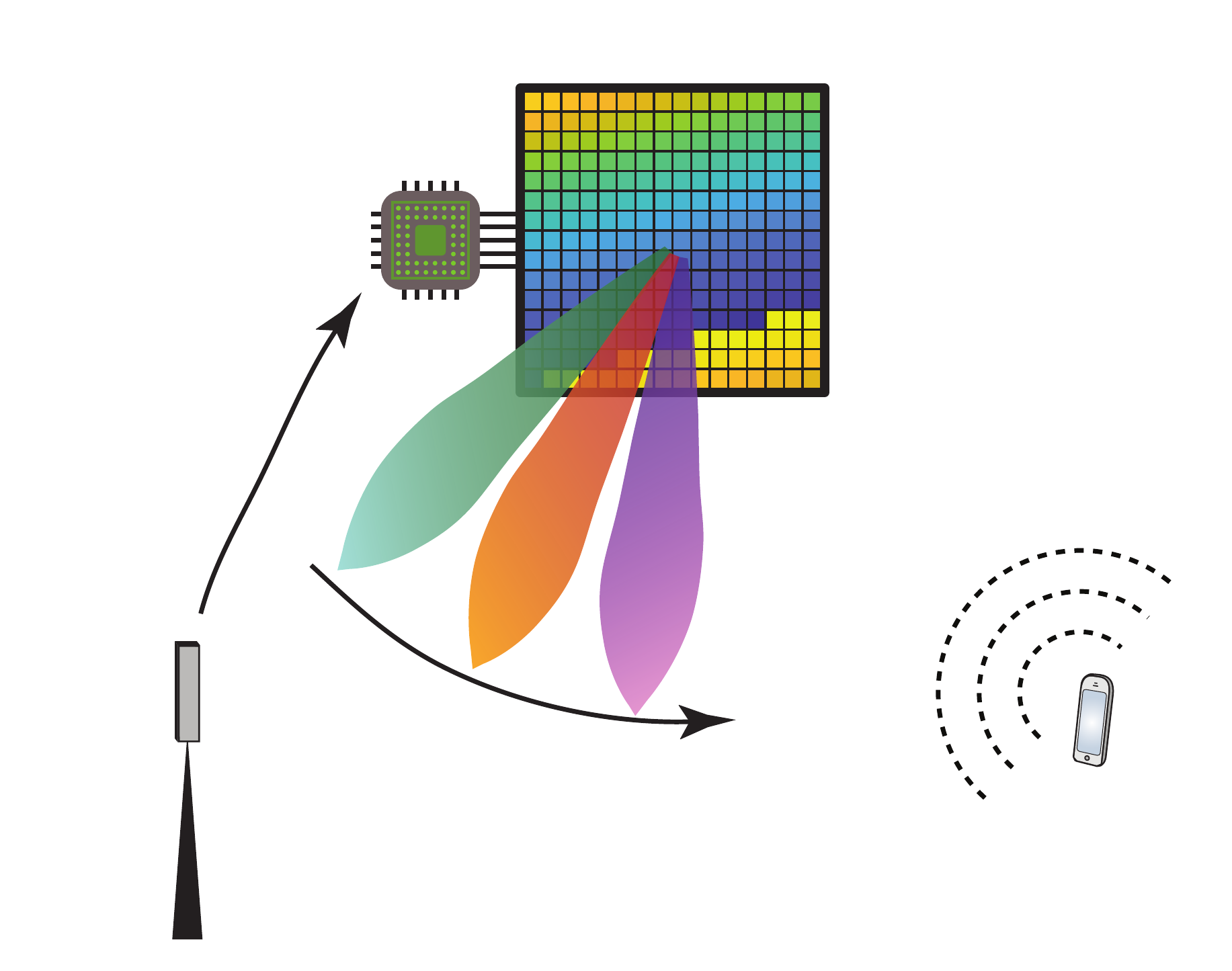}
\put(72,10){1) Repeated pilot}
\put(72,6){transmission}
\put(23,15){2) Switching between}
\put(23,11){different configurations}
\put(0,50){3) Feedback}
\put(0,46){of preferred}
\put(0,42){configuration}
\put(17,66){RIS controller}
\put(51.5,73){RIS}
\end{overpic} 
	\caption{One approach to configure the RIS is to transmit pilots that the RIS scatters using different configurations. The receiver feeds back a preferred configuration to the RIS.}
	\label{figureEstimation} 
\end{figure}

The literature contains initial approaches to tackle the problem. One approach is to transmit a pilot sequence repeatedly and measure the received signal when using different RIS configurations. For example, the elements can be turned on/off according to a pattern or the array geometry can be used to sweep through changes of the main reflection angle. 
At least $N$ reconfigurations must be tested in different time slots to excite all the channel dimensions. Only a concatenation of the channels to/from the RIS are observed and mutual coupling between RIS elements complicates the estimation. 
 This approach is illustrated in Fig.~\ref{figureEstimation} and requires a wireless control loop between the receiver and the RIS controller circuit with a capacity proportional to $N$. 
Even when CSI has been acquired, it is computationally complex to select appropriate time-delays, particularly in wideband channels \cite{Zheng2020}.
 To reduce complexity, adjacent RIS elements can be grouped to have the same configuration \cite{Zheng2020}, at the cost of a performance loss.

Another approach is to alter the passive nature of the RIS by having a few elements with receiver chains
\cite{Taha2019a}, which enables sensing and channel estimation directly at the RIS. 
The ability to extrapolate a few measurements to estimate the entire wideband channel requires spatially sparse channels with a known parametrization. This might be reasonable in mmWave or terahertz bands but further work on channel and hardware modeling is required. The sparseness can also make the channels flat over rather wide bandwidths. Learning-based and sparsity-based estimation algorithms were considered in \cite{Taha2019a,Alexandropoulos2020}. Even if the RIS has sensing capabilities, a control loop is needed to jointly select the RIS configuration and the beamforming at the transmitter/receiver.

Estimation algorithms can leverage special channel characteristics to reduce the pilot overhead. For instance, the channel between the BS and RIS is semi-static and common for all users, which makes the end-to-end channels correlated between users. An estimation algorithm exploiting this correlation was proposed in \cite{wang2019}. The BS-to-RIS channel can contain many coefficients if the BS has many antennas but since this channel is semi-static, it can be estimated less frequently than the RIS-to-user channel, which typically contains fewer coefficients since users  have fewer antennas.

There is no doubt that RIS can be used for fixed communication links, but mobile operation requires real-time channel estimation and reconfiguration, even in indoor use cases. 
A few millimeters of movement will change the channels in mmWave bands and above.
It remains to be demonstrated if any estimation protocol can enable real-time reconfigurability and under what mobility conditions. Since the array is passive, the RIS technology is potentially more energy-efficient than alternative technologies \cite{Basar2019a} but this remains to be demonstrated quantitatively. The RIS will require a power source for reconfigurability and wireless control channels.
It is likely that the control interface will consume most of the power at the RIS, so one cannot predict the total power consumption until the channel estimation and reconfigurability have been solved and validated.

\section*{Summary}

An RIS is a full-duplex transparent relay that synthesizes the scattering behavior of an arbitrarily shaped object. Since the RIS is not amplifying the signal, a larger surface area is required to achieve a given SNR than using conventional relays or multi-antenna transceivers. RIS-aided communication is an emerging research topic where the main open problems are to identify convincing use cases and designing practical protocols for reconfigurability.

\bibliographystyle{IEEEtran}

\section*{Biographies}

\textbf{Emil Bj\"ornson} is Associate Professor at Link\"oping
University, Sweden.
He has co-authored the textbooks \emph{Optimal Resource Allocation in
  Coordinated Multi-Cell Systems} (2013) and \emph{Massive MIMO
  Networks: Spectral, Energy, and Hardware Efficiency} (2017). He
received the 2018 IEEE Marconi Prize Paper Award in Wireless
Communications and the 2019 IEEE ComSoc Fred W. Ellersick
Prize.

\textbf{{\"O}zgecan {\"O}zdogan} is a Ph.D. student at Link\"oping
University, Sweden. She received her M.Sc. degree from Izmir Institute of Technology, Turkey, in 2017.

\textbf{Erik G. Larsson} is Professor at Link\"oping University,
Sweden, and Fellow of the IEEE.  He co-authored \emph{Fundamentals of
  Massive MIMO} (Cambridge University Press, 2016).  He received,
among others, the 2015 IEEE ComSoc Stephen O. Rice Prize in Communications
Theory, the 2017 IEEE ComSoc Leonard G. Abraham Prize, the 2018
IEEE ComSoc Best Tutorial Paper Award, and the 2019 IEEE ComSoc
Fred W. Ellersick Prize.  

\end{document}